\documentclass[aps,pre,twocolumn,groupedaddress,showpacs,floatfix]{revtex4}
\usepackage{graphicx}
\usepackage[latin1]{inputenc}
\usepackage{dcolumn}
\usepackage{bm}
\usepackage{amssymb}
\usepackage{amsmath}
\usepackage{epsfig}
\def\LM#1#2{\left|\begin{array}{l}{#1}\\[1ex]{#2}\end{array}\right.}

\begin{document}
\title{The target problem with evanescent subdiffusive traps}
\author{S. B. Yuste$^{1}$, J. J. Ruiz-Lorenzo$^{1}$, and
Katja Lindenberg$^{2}$}
\affiliation{$^{(1)}$ Departamento de F\'{\i}sica, Universidad de
Extremadura, E-06071 Badajoz, Spain\\
$^{(2)}$ Department of Chemistry and Biochemistry 0340, and Institute for
Nonlinear Science,
University of California San Diego, 9500 Gilman Drive, La Jolla, CA
92093-0340, USA}
\begin{abstract}
We calculate the survival probability of a stationary target in one
dimension surrounded by diffusive or subdiffusive traps of
time-dependent density.  The survival probability of a target in the
presence of traps of constant density
is known to go to zero as a stretched exponential whose specific
power is determined by the exponent that characterizes the motion of
the traps. A density of traps that grows in time always leads to an
asymptotically vanishing survival probability.  
Trap evanescence leads to a survival probability of the
target that may be go to zero or to a finite
value indicating a probability of eternal survival, depending on the
way in which the traps disappear with time.
\end{abstract}

\maketitle
\section{INTRODUCTION}
\label{sec:intro}
The traditional {\em trapping} problem involves diffusive (Brownian)
particles ($A$) that wander in a medium doped with static traps
($B$) and disappear when they
meet~\cite{Hughes,Weiss,ShDba,AvrahamHavlinDifuReacBook}. In the
traditional {\em target}
problem~\cite{Klafter,BluKlaZuOptical,SungJCP02}, on the
other hand, one has static
$A$ particles and mobile traps.  Both of these problems are described by
the ``reaction" $A+B \rightarrow B$, but in one case the $A$'s
move and the $B$'s stand still, while in the other the $B$'s move while
the $A$'s are stationary.  Both of these
problems have a long and active history in the literature.  They
not only represent experimentally observable phenomena, but they have
served as a testbed for theoretical and numerical studies
and as a starting point for the formulation of models for more complex
systems that have only recently
been successfully solved analytically.  For example, the
survival probability of an $A$ particle in a medium of $B$ particles
when {\em both} species are diffusive, first investigated numerically in
the seminal work of Toussaint and Wilczek~\cite{ToussaintWilczekPRL},
was only partially solved
analytically~\cite{BramsonLebowitz1,BramsonLebowitz2}
until the recent full (asymptotic) solution in one
dimension~\cite{BrayBlythePRLPRE,OshaninEtAlPRE,MoreauEtAlCondMat,BrayMajumBlythePRE}.
These results have also recently been generalized to
subdiffusive species~\cite{ourPRE}.
The survival probability of $A$ particles in the
reactions $A+A \rightarrow A$ and $A+A \rightarrow 0$ in one dimension
when $A$ is mobile is also of relatively recent vintage in the history
of such analytic
solutions~\cite{AvrahamHavlinDifuReacBook,Habibetc,YusteKatja}.

The purpose of this paper is to extend the one-dimensional target
problem calculations for both diffusive and subdiffusive traps to the
case of traps that themselves disappear in time according to some
survival probability function of their own (e.g., exponential or power
law). The decay of the moving traps with time of course increases the
survival probability of the stationary target, and the interesting
questions concern the interplay of the time dependences of the movement
and decay of the traps.  A related problem was considered in~\cite{Wio},
where diffusive particles $A$ and traps $B$ and $C$ undergoing the
explicit reactions (a) $A+B\to B$, $B+C\to C$, and (b) $A+B\to B$,
$B+C\to 0$ were considered using entirely different methods.
Our methods are equally applicable to trap densities that increase with
time, but this problem is less interesting because it necessarily leads
to the eventual demise of the target.

A common characterization of the diffusive
motion of a particle is through its mean square displacement
for large $t$,
\begin{equation}
\left< x^2(t)\right> \sim \frac{2K_\gamma}{\Gamma(1+\gamma)}
t^\gamma . \label{meansquaredispl}
\end{equation}
Here $K_\gamma$ is the (generalized) diffusion constant, and
$\gamma$ is the exponent that characterizes normal ($\gamma=1$) or
anomalous ($\gamma\neq 1$) diffusion.  In particular,
the process is sudiffusive when $0<\gamma<1$. Subiffusive processes are
ubiquitous in
nature~\cite{MetKlaPhysRep,BouchaudPhysRep90,Kosztolowicz,SubdifuRandPot1,SubdifuRandPot,KantorCM},
and are particularly useful for understanding transport in complex
systems~\cite{ShDba,BouchaudJPI}.

The problem considered in this paper is a special case of
a broad class of reaction-\emph{subdiffusion} processes that have been
studied over the past decades. One approach that has been used to study
these processes is based on the continuous time random
walk (CTRW) theory with waiting-time distributions between steps that
have broad long-time tails and consequently infinite moments,
$\psi(t)\sim t^{-1-\gamma}$ for $t\to \infty$ with $0<\gamma<1$.
Another approach is based on the fractional diffusion equation,
which describes the evolution of the probability density $P(x,t)$
of finding the particle at position $x$ at time $t$ by means of
the fractional partial differential equation (in one
dimension)~\cite{SungJCP02,MetKlaPhysRep,SchWysJMP,SekiJCP1s03,YusteAcedoSubTrap,YusAceLinSubFront}
\begin{equation}
\frac{\partial }{\partial t} P(x,t)= K_\gamma
~_{0}\,D_{t}^{1-\gamma } \frac{\partial^2}{\partial x^2} P(x,t),
\label{Pfracdifu}
\end{equation}
where $~_{0}\,D_{t}^{1-\gamma } $ is the Riemann-Liouville
operator,
\begin{equation}
~_{0}\,D_{t}^{1-\gamma } P(x,t)=\frac{1}{\Gamma(\gamma)}
\frac{\partial}{\partial t} \int_0^t dt'
\frac{P(x,t')}{(t-t')^{1-\gamma}}.
\end{equation}

In this paper we study the one-dimensional target problem
for a static particle $A$ subject to attack by
diffusive or subdiffusive traps $B$ that may die before reaching the
target $A$~\cite{Wio}.  For this purpose, we generalize the
ideas of Bray and Blythe~\cite{BrayBlythePRLPRE}, and of our own
work~\cite{ourPRE} based on a fractional diffusion equation approach.
While recent work shows that a
simple generalization of
reaction-diffusion to reaction-subdiffusion equations in which the
reaction and subdiffusion terms are assumed to enter
additively is not valid in
some cases~\cite{newigor}, this is not a difficulty in
our particular application.  The difficulties do not arise when the
reaction process can be translated into a static boundary value problem,
which is the case for the target (as well as the trapping) problem  \cite{YusteAcedoSubTrap}.

In some cases, asymptotic anomalous diffusion behavior can be found from
corresponding results for normal diffusion via the simple replacement
of $t$ by $t^\gamma$. This can be understood from
a CTRW perpective because
the average number of jumps $n$ made by a subdiffusive walker up to
time $t$
scales as $\langle n \rangle \sim t^\gamma$ and, in many instances the
number of jumps is the relevant factor that explains the
behavior of the system.
The simple replacement result is evidence
of ``subordination'' (see Secs. 5 and 7.2 of~\cite{BluKlaZuOptical}).
However, there are other instances where
the behavior of subdiffusive systems cannot be found in this way.
A simple example is the survival probability of subdiffusive
particles in the trapping problem (see Sec. 5
of~\cite{BluKlaZuOptical}).  In particular, for
systems where competing processes (motion toward target and death)
occur according to different temporal rules,
such a replacement becomes ambiguous.
This is the case for the problem considered here.

While our analytic results are based on the fractional diffusion
equation formalism, our numerical simulations are based on a CTRW
algorithm.  These two renditions of the problem are expected to differ if
trapping events are likely in a small number of steps, that is, if the
initial density of traps is too high.  On the other hand, if the initial
trap density is too low, then the simulations to produce valid
statistics would take inordinately long because trapping events are rare
and because the system has to be sufficiently large to include many
particles.  We note this as a caveat for our subsequent comparisons.

In Sec.~\ref{survival1} we present an integral equation for the survival
probability, which we reduce to quadrature in
Sec.~\ref{survival2}.  The resulting integral is explicitly evaluated
for exponentially decaying trap densities (including a stretched
exponential decay), as well as for trap densities that decay as a power
law.  Not surprisingly, we find that a sufficiently rapid decay of the
trap density leads to a finite asymptotic survival probability of the
target.  Comparisons of our results with numerical simulations
are also shown in this section.
A summary and some conclusions are presented in Sec.~\ref{conclusions}.

\section{INTEGRAL EQUATION FOR THE SURVIVAL PROBABILITY}
\label{survival1}
We consider a finite
interval $L$ containing $N=\rho L$ mobile traps $B$ of constant density
$\rho$ initially distributed
at random, and a single immobile $A$ particle at the origin.
Following the approach of
Bray et al.~\cite{BrayMajumBlythePRE} for diffusive traps and
our generalization of this approach to the subdiffusive
case~\cite{ourPRE}, we write the survival probability of $A$ as
$P(t)= \exp\{-\mu_0(t)\}$, where $\mu_0(t)$ is to
be determined.  To find this function, one calculates
in two ways the probability density to find
a $B$ particle at the origin at time $t$,
\begin{equation}
\rho =  \int_0^t dt' \dot{\mu}_0(t') G(t-t').
\label{fundamental}
\end{equation}
That the left side is this probability density is obvious.
On the right side one has the renewal theory expression where
$\dot{\mu}_0(t')dt'= (-\dot{P}/P)dt'$ is the probability
that a $B$ particle intersected $A$ in the time interval $(t',t'+dt')$
for the first time, and the propagator $G(t-t')$ is
the probability density for this particular $B$ to be at the origin at
time $t$. In one dimension it is given
by~\cite{MathaiSaxena,MetzlerKlafter}
\begin{eqnarray}
G(t) &=& \frac{1}{\sqrt{4\pi K_\gamma t^\gamma}} H_{1,2}^{2,0}\left[0
\LM{(1-\gamma/2 ,\gamma/2)}{(0,1),(1/2,1)} \right]
\nonumber\\
&=&
\frac{1}{\sqrt{4K_\gamma t^\gamma} \Gamma\left(1-\frac{\gamma}{2}\right)},
\end{eqnarray}
where $H_{1,2}^{2,0}$ is Fox's
$H$-function, whose value at the
given arguments we have used to write the last equality.
In a different context than the target problem,
Bray et al.~\cite{BrayMajumBlythePRE} generalized their approach to
a time-dependent density $\rho(t)$ of $B$.
They argue that in place of Eq.~(\ref{fundamental}) one now has
\begin{equation}
\label{ecuIntegral0}
\rho(t) = \int_0^t dt' \frac{\rho(t)}{\rho(t')} \dot{\mu}_0 G(t-t') ,
\end{equation}
that is,
\begin{equation}
\label{ecuIntegral1}
1 = \int_0^t dt' \frac{\dot{\mu_0}}{\rho(t')} G(t-t').
\end{equation}
This is the basic equation to be considered in this paper.

\section{THE SURVIVAL PROBABILITY}
\label{survival2}
To calculate the survival probability of particle $A$,
we rewrite Eq.~(\ref{ecuIntegral1}) explicitly as
\begin{equation}
\label{ecuAbel}
\sqrt{4K_\gamma} = \frac{1}{ \Gamma(1-\gamma/2)}\int_0^t dt'
\frac{\Omega(t')}{(t-t')^{\gamma/2}} ,
\end{equation}
where we have multiplied both sides by $\sqrt{4K_\gamma}$, and where
we have introduced
\begin{equation}\label{}
    \Omega(t)=\frac{\dot \mu_0(t)}{\rho(t)}.
\end{equation}
Equation~(\ref{ecuAbel}) is an equation of Abel of the first
kind~\cite{Tricomi,GorenfloMainardi97},
\begin{equation}
\label{ecuAbel1}
f(t) = \frac{1}{ \Gamma(1-\gamma/2)}\int_0^t dt'
\frac{\Omega(t')}{(t-t')^{\gamma/2}} ,
\end{equation}
with $f(t)=\sqrt{4K_\gamma}$. The solution of this classic
equation is well known (see
Sec.~12 in \cite{Tricomi} or
Eqs.~(2.5a) and (2.5b) in ~\cite{GorenfloMainardi97}),
\begin{align}\label{}
    \Omega(t)&=~_{0}D_{t}^{1-\gamma/2} f(t)  \nonumber \\
&=\frac{1}{\Gamma(\gamma/2)}\int_0^t dt' \frac{\dot
f(t')}{(t-t')^{1-\gamma/2}} +f(0^+)
\frac{t^{\gamma/2-1}}{\Gamma(\gamma/2)} .
\end{align}
Here, as earlier, $~_{0}D_{t}^{1-\gamma/2}$ is the
Riemann-Liouville fractional derivative.
In our case $f(t)=\sqrt{4K_\gamma}$ is constant, so that
\begin{equation}\label{solOmega}
    \Omega(t)=\frac{\dot \mu_0(t)}{\rho(t)}=
\frac{\sqrt{4K_\gamma}}{\Gamma(\gamma/2)}\, t^{\gamma/2-1}.
\end{equation}
It then follows that
\begin{equation}\label{musolInte}
    \mu_0(t)=\frac{\sqrt{4K_\gamma}}{\Gamma(\gamma/2)} \int_0^t dz
\rho(z) z^{\gamma/2-1},
\end{equation}
which provides a general solution to our problem for any $\rho(t)$.
While Eq.~(\ref{musolInte}) applies to trap densities that grow or that
decrease or even oscillate in time, the case of evanescent traps is the
more interesting and the one we choose to focus on.

\begin{figure}
\begin{center}
\includegraphics[width=1.0\columnwidth]{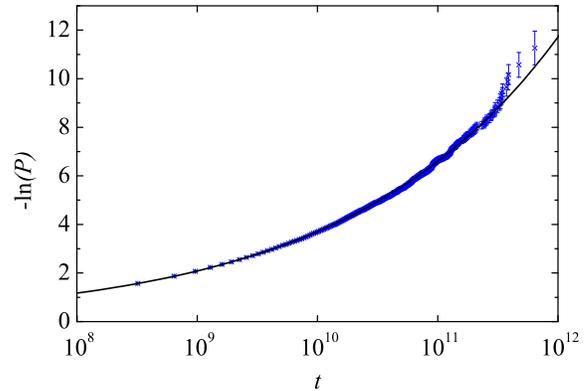}
\caption{ (Color online)
$\mu_0(t)=-\ln P(t)$ vs $t$ for non-evanescent traps as given in
Eq.~(\ref{nonevan}) (solid line) and simulations (symbols along with error
bars). Parameter values are
$\gamma=1/2$, $\rho_0=0.01$, $K_\gamma=1/(2\sqrt{\pi})$.
\label{fig:mu0_ga0.5_rho0.01_a0}}
\end{center}
\end{figure}

As a reminder, we note that for traps of a constant density
$\rho(t)=\rho_0$ the survival probability is given by
\begin{equation}
P(t) = \exp \left( \frac{\sqrt{4K_\gamma} \rho_0
t^{\gamma/2}}{\Gamma(1+\gamma/2)}\right).
\label{nonevan}
\end{equation}
As a benchmark, we show in
Fig.~\ref{fig:mu0_ga0.5_rho0.01_a0} a typical comparison of this
result with simulation results. The agreement is clearly good, although
a lower initial density run for a longer time would
lead to even better agreement.

\subsection{Exponentially decaying trap density}
\label{expdec}
Suppose that the traps have a finite lifetime $\tau$ and decay
exponentially, as in a unimolecular reaction, $\rho(t) =
\rho_0\exp(-t/\tau)$.
The integral in Eq.~(\ref{musolInte}) immediately
leads to the solution
\begin{equation}\label{mu0}
    \mu_0(t)=\ell_\gamma \rho_0
\left(1-\frac{\Gamma(\gamma/2,t/\tau)}{\Gamma(\gamma/2)}\right),
\end{equation}
where $\Gamma(b,x)$ is an incomplete Gamma function, and
\begin{equation}
\ell_\gamma \equiv(4K_\gamma \tau^\gamma)^{1/2}.
\label{ell}
\end{equation}
When $\gamma=1$, i.e., when the traps are diffusive, this
reduces to
\begin{equation}
\mu_0(t)=\ell_1 \rho_0\text{erf}(\sqrt{t/\tau}).
\label{forfig2}
\end{equation}
For arbitrary $\gamma<1$, the survival probability of the target in the
presence of the subdiffusive traps with finite lifetime thus is
\begin{equation}\label{}
  P(t)=\exp\left[-\ell_\gamma \rho_0
\left(1-\frac{\Gamma(\gamma/2,t/\tau)}{\Gamma(\gamma/2)}\right)\right].
\label{forfig3}
\end{equation}

The interesting result here is that the funtion $\mu_0(t)$
\emph{goes to the constant} $\mu_0(\infty) = \ell_\gamma \rho_0$
and not to infinity as $t\to\infty$. Therefore
the survival probability does not vanish with increasing time,
\begin{equation}\label{pas}
P(t) \to \exp\left[-\ell_\gamma \rho_0
\left( 1-\frac
{e^{-t/\tau}}{\Gamma(\gamma/2)(t/\tau)^{1-\gamma/2}}\right)\right].
\end{equation}
We note that $\ell_\gamma$
is a characteristic distance that measures the root mean square
displacement of the traps during their decay time $\tau$.  Therefore
$\mu_0(\infty)=\ell_\gamma \rho_0$ is the ratio of this average displacement
to the average initial distance $\rho_0^{-1}$ between traps.
This finite asymptotic survival probability,
$P(\infty)=\exp\left(-\ell_\gamma \rho_0\right)$, displays reasonable
qualitative features: it increases with decreasing trap lifetime
$\tau$, and it decreases with increasing initial trap density $\rho_0$.
That there is a finite asymptotic survival probability reflects the fact
that if the traps disappear sufficiently rapidly (which they do if they
disappear exponentially while the traps move diffusively or
subdiffusively), then many traps disappear before they can reach the
particle, and there is a finite probability that the particle remains
forever ``safe."

\begin{figure}
\begin{center}
\includegraphics[width=1.0\columnwidth]{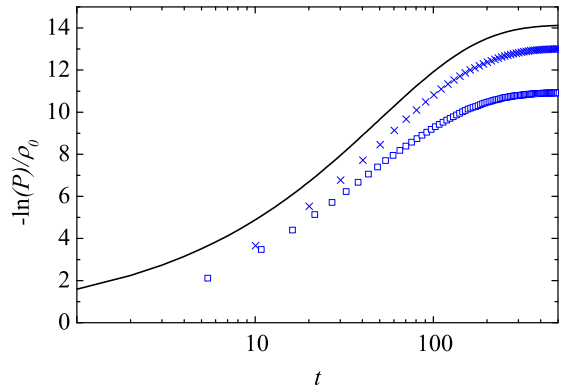}
\caption{(Color online) $\mu_0(t)/\rho_0=-\ln P(t)/\rho_0$ vs $t$ for
exponentially evanescent traps. Solid line: Eq.~(\ref{forfig2}).
Squares: simulation results for a high initial density $\rho_0=0.1$.
$X$'s: simulation results for a lower initial density $\rho_0=0.01$.
Other parameter values are
$\gamma=1$, $\tau=100$, and $K_1=D = 1/2$. Asymptotic value:
$\mu_0(\infty)=\ell_1\rho_0=\sqrt{2}$.
\label{fig:mu0_rho0.1_a0.01}}
\end{center}
\end{figure}

\begin{figure}
\begin{center}
\includegraphics[width=1.0\columnwidth]{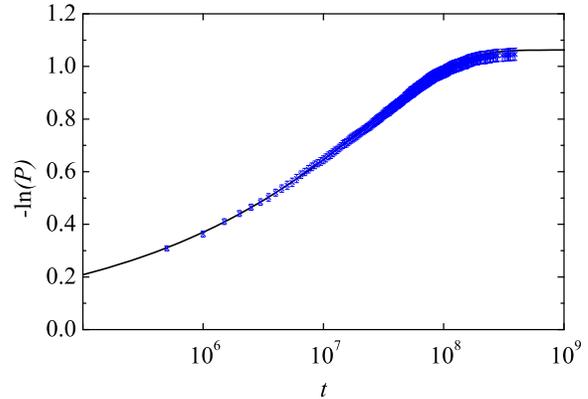}
\caption{(Color online) $\mu_0(t)=-\ln P(t)$ vs $t$ for
exponentially evanescent traps
of a lower initial density $\rho_0=0.01$. Solid line:
Eq.~(\ref{forfig3}). Symbols: simulation results (there are error bars
on the symbols but they are too small to see clearly).
Other parameter values are
$\gamma=1/2$, $\tau=10^8$, and $K_\gamma= 1/(2\sqrt{\pi})$.
Asymptotic value: $\mu_0(\infty)=\ell_{1/2}\rho_0=1.06225$.
\label{fig:mu0_ga0.5_rho0.01_a1e-8}}
\end{center}
\end{figure}

The next two figures show the comparison of simulation results with
our analytic outcome.  First, in Fig.~\ref{fig:mu0_rho0.1_a0.01}
we illustrate our earlier
caveat, that agreement cannot be expected if the initial density of
traps is too high and the extinction rate of the traps is large, and
that the agreement improves with lower initial density. 
The disagreement is clear and can be traced exactly
to the early time trapping events that cumulatively affect the survival
probability.  Fig.~\ref{fig:mu0_ga0.5_rho0.01_a1e-8} shows typical
results for the lower initial
density of traps and a more slowly decaying trap density, where
the agreement between analytic results and
simulations is clearly very good.

Finally, it is straightforward to extend the results of this section to
trap densities that decay as a stretched exponential, $\rho(t) = \rho_0
\exp [-(t/\tau)^\alpha]$. The integral \eqref{musolInte} is still
straightforward and gives
\begin{equation}\label{mu0s}
    \mu_0(t)=\frac{\ell_\gamma \rho_0\Gamma(\gamma/2\alpha)}
{\alpha \Gamma(\gamma/2)}
\left(1-\frac{\Gamma(\gamma/2\alpha,(t/\tau)^\alpha)}{\Gamma(\gamma/2\alpha)}
\right),
\end{equation}
which reduces to Eq.~(\ref{mu0}) when $\alpha=1$.  The asymptotic
finite survival probability then is
\begin{equation}\label{pass}
P(t)\to \exp\left[-\ell_\gamma \rho_0
\frac{\Gamma(\gamma/2\alpha)}{\alpha\Gamma(\gamma/2)}
\left( 1-\frac
{e^{(-t/\tau)^\alpha}}{\Gamma(\gamma/2\alpha)(t/\tau)^{\alpha(1-\gamma/2)}}
\right)\right].
\end{equation}
An interesting interplay of $\gamma$ and $\alpha$ should be noted:
there are values of $\alpha$ and $\gamma$ for which the survival
probability of the target when the trap density decays as a
stretched exponential ($\alpha<1$) is
actually greater than with an exponential decay ($\alpha=1$).  This
seemingly counterintuitive behavior is connected with the
reversal of time inequalities, i.e., with the fact that
$(t/\tau)^\alpha$ is greater (smaller) than $(t/\tau)$ when
$t$ is smaller (greater) than $\tau$.

\subsection{Power law decaying trap density}
\label{powerlawdec}

Suppose now that the trap density decays as a power law as might happen,
for instance, if there is a process of trap-trap annihilation.  The
trap density at long times then decreases as $\rho(t)\sim t^{-\beta}$
and it is to be expected that the target survival probability (and, in
particular, whether it is asymptotically vanishing or finite) depends
sensitively on the relation between the exponents $\beta$ and $\gamma$.
We expect that for sufficiently large $\beta$ the target will again
have a finite probability of surviving forever.

To find a closed expression for the survival probability we need to
specify $\rho(t)$ for all times, not just asymptotically, and we choose
\begin{equation}\label{}
\rho(t)=\frac{\rho_0}{(1+t/\tau)^\beta}.
\end{equation}
With this form, the integral~(\ref{mu0}) can be carried out exactly, to
give
\begin{equation}
\mu_0(t) = \frac{\ell_\gamma \rho_0}{\Gamma(\gamma/2)}
B_{\frac{t}{(\tau+t)}}(\gamma/2,\beta-\gamma/2)
\label{general}
\end{equation}
for all $\beta$, where $B$ is the incomplete Beta function~\cite{web,Abramo}
\begin{equation}
B_\alpha(z,w) = \int_0^\alpha dt \; t^{z-1} (1-t)^{w-1} \quad
\text{with } \Re(z)>0.
\label{ibf}
\end{equation}

Equation~\eqref{meansquaredispl} tells us that
the typical length explored by a
(living) trap grows with time as $\langle x^2(t)\rangle^{1/2}
\sim t^{\gamma/2}$. On the other hand, the mean
distance between traps grows as $\rho^{-1} \sim t^\beta$. It thus stands
to reason that the asymptotic survival probability depends
sensitively on the relative magnitudes of $\beta$ and $\gamma/2$.
To present more explicit results in this long-time regime we
distinguish three cases.

\begin{figure}
\begin{center}
\includegraphics[width=1.0\columnwidth]{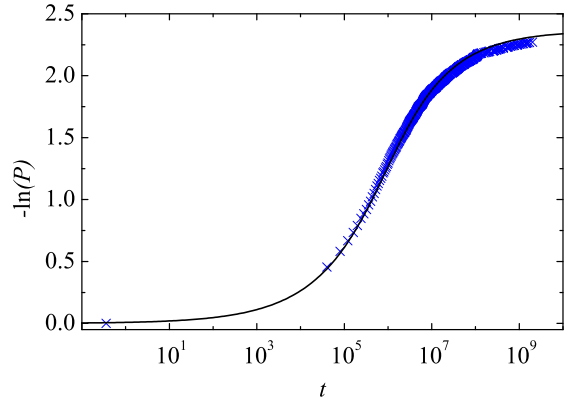}
\caption{(Color online) $\mu_0(t)=-\ln P(t)$ vs $t$ for power law
evanescent traps
with $\beta>\gamma/2$ (``Case 1"). Parameter values are
$\gamma=0.75$, $\beta=0.8$, $\rho_0=0.01$, $\tau=10^6$,
and $K_\gamma= 1/(2\sqrt{\pi})$.
Asymptotic value: $\mu_0(\infty)=\ell_{3/4}\rho_0=2.36549$.
Solid line: Eq.~(\ref{general}). Symbols: simulation results.
\label{fig:mu_ga0.75_b1e-6beta0.8_L1e4_rho1e-2}}
\end{center}
\end{figure}

{\bf Case 1: $\beta > \gamma/2$}. In this case Eq.~(\ref{general})
can be written as
\begin{equation}
    \mu_0(t)= \frac{\ell_\gamma \rho_0}{\Gamma(\gamma/2)}
B(\gamma/2,\beta-\gamma/2) I_{\frac{t}{(\tau+t)}}(\gamma/2,\beta-\gamma/2).
\end{equation}
Here $B(z,w)$ is the Beta function (where the requirement $\Re(z)>0$
and $\Re(w)>0$ places us in the ``Case 1" regime), and
$I_{x}(z,w)$ is the regularized incomplete Beta function as
defined in Sec.~6.6.2 (pg.~263) of~\cite{Abramo}. Using the
property~6.6.3 in~\cite{Abramo} we can set
$I_{x}(a,b)=1- I_{1-x}(b,a)$, and applying the
relation 26.5.5 in~\cite{Abramo} we can then write the asymptotic result
\begin{equation}\label{}
I_{\frac{t}{(\tau+t)}}(\gamma/2,\beta-\gamma/2)=1-\frac{(t/\tau)^
{\gamma/2-\beta}}{(\beta-\gamma/2)B(\beta-\gamma/2,\gamma/2)}+\ldots .
\end{equation}
Consequently, recognizing the relation between
the Beta function and the Gamma function, as $t \to\infty$ we arrive at
the asymptotic result
\begin{equation}\label{}
\mu_0(t) \to
\ell_\gamma \rho_0
\left(\frac{\Gamma(\beta-\gamma/2)}{\Gamma(\beta)}-
\frac{(t/\tau)^{\gamma/2-\beta}}{(\beta-\gamma/2)
\Gamma(\gamma/2)}+\ldots\right).
\end{equation}
The survival probability thus approaches
(via a power law decay of the exponent) the finite asymptotic value
\begin{equation}
P(t\to\infty) = \exp \left( - \ell_\gamma \rho_0 \,
\frac{\Gamma(\beta-\gamma/2)}{\Gamma(\beta)}\right).
\label{case1}
\end{equation}
Figure~\ref{fig:mu_ga0.75_b1e-6beta0.8_L1e4_rho1e-2}
illustrates this result along with numerical simulations for comparison.

\begin{figure}
\begin{center}
\includegraphics[width=1.0\columnwidth]{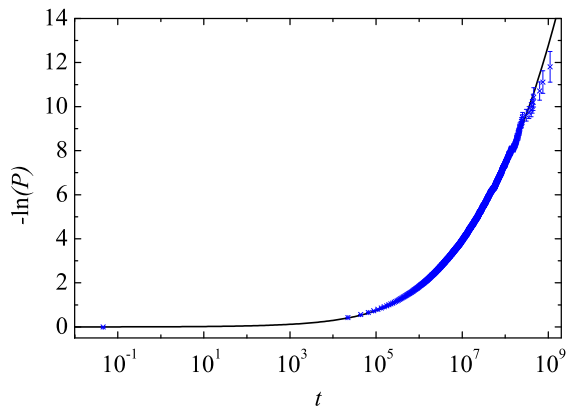}
\caption{(Color online)
$\mu_0(t)=-\ln P(t)$ vs $t$ for power law evanescent traps
with $\beta<\gamma/2$ (``Case 2"). Parameter values are
$\gamma=0.8$, $\beta=0.2$, $\rho_0=0.01$, $\tau=10^6$,
and $K_\gamma= 1/(2\sqrt{\pi})$.  Solid line: Eq.~(\ref{general}).
Symbols: simulation results along with error bars.
\label{fig:mu_ga0.8_b1e-6_beta0.2_L1e4_rho0.01}}
\end{center}
\end{figure}

{\bf Case 2: $\beta < \gamma/2$}.
In this case the integrand in Eq.~\eqref{musolInte} goes to zero
more slowly than $1/t$ for $t \to \infty$, so that a simple
asymptotic analysis of the integral \eqref{musolInte}
readily establishes that $\mu_0(t)$ goes to infinity
with increasing time as
\begin{equation}
\mu_0(t) \to  \frac{\ell_\gamma \rho_0}{(\gamma/2-\beta)\Gamma(\gamma/2)}
\left( \frac{t}{\tau}\right)^{\gamma/2-\beta} + \ldots,
\end{equation}
so that the survival probability vanishes at long times as a stretched
exponential,
\begin{equation}
\label{case2}
P(t) \to \exp \left(-\frac{\ell_\gamma}{(\gamma/2 -
\beta)\Gamma(\gamma/2)}
(t/\tau)^{\gamma/2-\beta}\right).
\end{equation}
Analytic and simulation results for this case are shown in
Fig.~\ref{fig:mu_ga0.8_b1e-6_beta0.2_L1e4_rho0.01}.

\begin{figure}
\begin{center}
\includegraphics[width=1.0\columnwidth]{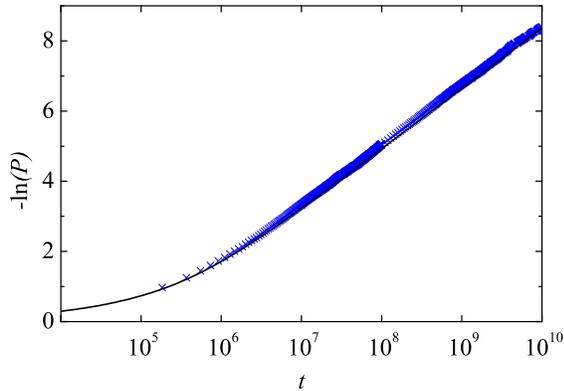}
\caption{(Color online)
$\mu_0(t)=-\ln P(t)$ vs $t$ for power law evanescent traps
with $\beta=\gamma/2$ (``Case 3"). Parameter values are
$\gamma=0.8$, $\beta=0.4$, $\rho_0=0.01$, $\tau=10^6$,
and $K_\gamma= 1/(2\sqrt{\pi})$.
Solid line: Eq.~(\ref{general}). Symbols: simulation results.
\label{fig:mu_ga0.8_b1e-6_beta0.4_L1e4_rho0.01}}
\end{center}
\end{figure}

{\bf Case 3: $\beta = \gamma/2$}.
This is the marginal case, and the incomplete Beta function~(\ref{ibf})
can be rewritten as a hypergeometric function,
\begin{align}
\mu_0(t) &= \frac{ \ell_\gamma \rho_0 (t/\tau)^\beta}{\beta
\Gamma(\gamma/2)} \; _2F_1(\beta,\beta, 1+\beta, -t/\tau) \nonumber\\
&= \frac{\ell_\gamma \rho_0}{\Gamma(\gamma/2)} \ln (t/\tau) +\ldots
\quad \text{as  } t\to\infty.
\end{align}
The survival probability thus decays as an inverse power,
\begin{equation}
P(t \to \infty) \to (t/\tau)^{-\ell_\gamma \rho_0/\Gamma(\gamma/2)}.
\label{marginal}
\end{equation}
Results for the marginal case are shown in
Fig.~\ref{fig:mu_ga0.8_b1e-6_beta0.4_L1e4_rho0.01}.

\section{Conclusions}
\label{conclusions}

We have calculated the survival probability of a stationary target in a
one-dimensional system in which diffusive or subdiffusive traps that
eliminate the target upon encounter themselves disappear according to a
survival probability.  The root mean square displacement of the traps
grows with time as $t^{\gamma/2}$, that is, diffusively when $\gamma=1$
and subdiffusively when $\gamma<1$.
The survival probability of the target depends
sensitively on the interplay of two temporal events, namely, the motion
of the traps as characterized by the exponent $\gamma$
and their disappearance.  When the motion of the traps is
diffusive or subdiffusive and the traps do not decay in time, the
survival probability goes to zero as a stretched exponential,
Eq.~(\ref{nonevan}).  When the traps undergo exponential decay or
stretched exponential decay, the target has an asympototic safety
margin, that is, a finite probability of surviving forever, cf.
Eqs.~(\ref{pas}) and (\ref{pass}).  When the
traps are diffusive or subdiffusive and disappear according to a power
law survival probability $\sim t^{-\beta}$, the survival of the target
depends sensitively on the relation between $\gamma$ and $\beta$.
If the traps move sufficiently rapidly relative
to their disappearance, that is, if $\gamma/2 >\beta$, the target is
trapped with certainty at long
times, its survival probability going to zero again as a stretched
exponential, cf. Eq.~(\ref{case2}).
If the traps move slowly, $\gamma/2 <\beta$, then the target
has a chance of eternal survival, cf. Eq.~(\ref{case1}).
At the critical relation
$\gamma/2=\beta$ the survival probability goes to zero as an inverse
power of time, cf. Eq.~(\ref{marginal}).  If in fact the trap density
increases with time, the survival probability of the target necessarily
vanishes asymptotically.

In this paper we have calculated the survival probability of a
target particle in the presence of evanescent subdiffusive traps of
given time-dependent density. We could equally consider the inverse
problem, namely, that of finding the time dependence of the density of
traps to obtain a particular survival probability function.
For this purpose we need only ``invert" Eq.~(\ref{solOmega}),
\begin{equation}
\rho(t) = - \frac{\Gamma(\gamma/2)}{\sqrt{4K_\gamma}} t^{1-\gamma/2}
\frac{\dot P(t)}{P(t)}.
\end{equation}
An exponentially decaying survival probability of the form
$P(t)=e^{-t/\tau}$ requires a density that decays as $\rho(t)\sim
t^{1-\gamma/2}$.  This is included in and consistent with Case 2 in
Sec.~\ref{powerlawdec} with $\gamma/2-\beta=1$.  Similarly, for an
inverse power decay of the form $P(t)\sim (t/\tau)^{-1}$ we require
that $\rho(t)\sim t^{-\gamma/2}$ consistent with Case 3 in the same
section.

This work has focused on the survival probability of a stationary
target.  The survival probability of a moving target, diffusive or
subdiffusive, surrounded by non-evanescent diffusive or subdiffusive traps
has been considered recently in a number of
papers~\cite{BrayBlythePRLPRE,ourPRE}.
Extension of our work with evanescent traps to the case of a diffusive
or subdiffusive target is in progress~\cite{nextone}.

\begin{acknowledgments}
The research of S.B.Y. and J.J.R.L has been supported by
the Ministerio de Educaci\'on y Ciencia (Spain) through grant No.\
FIS2004-01399 (partially financed by FEDER funds) and by the
European Community's Human Potential Programme under contract
HPRN-CT-2002-00307, DYGLAGEMEM. K.L. is supported in part by the
National Science Foundation under grant PHY-0354937.
\end{acknowledgments}

\end{document}